\documentclass[11pt]{article}
\usepackage{graphicx}
\usepackage{amsmath}
\usepackage{float}
\usepackage[colorlinks=true,urlcolor=blue,linkcolor=blue]{hyperref}
\usepackage[english]{babel}
\usepackage{color}

\def\RR{{\rm I\kern-.17em R}}
\usepackage{amsmath}
\usepackage{amssymb}
\usepackage{amsfonts}
\def\RR{{\rm I\kern-.17em R}}
\usepackage{color}
\usepackage{graphicx}
\usepackage{float}
\usepackage{graphics}
\newcommand{\be}{\begin{eqnarray}}

\newcommand{\ee}{\end{eqnarray}}
\newcommand{\dep}{\partial}

\begin{document}

\title{Elastic thick shells in General Relativity}
\author{Irene Brito$^\ast$, J. Carot$^\natural$, E.G.L.R. Vaz$^\ast$\\\\
$^\ast$Centro de Matem\'atica, Departamento de Matem\'atica,\\
Escola de Ci\^encias, Universidade do Minho, 4800-058 Guimar\~aes, Portugal\\
$^\natural$Departament de F\'isica, Universitat de les Illes
Balears,\\ Cra Valldemossa pk 7.5, E-07122 Palma de Mallorca, Spain
}

\maketitle
\begin{abstract}
It is shown that exact spherically symmetric solutions to Einstein's Field Equations exist such that,
over an open region of the spacetime, they are singularity free, satisfy the dominant energy condition,
represent elastic matter with a well defined constitutive function, and are such that elastic perturbations
propagate causally.

Two toy-models are then built up in which a thick elastic, spherically symmetric shell with the above properties,
separates two Robertson-Walker regions corresponding to different values of the curvature $k$.
The junction conditions (continuity of the first and second fundamental forms) are shown to be
exactly satisfied across the corresponding matching spherical surfaces.

\end{abstract}

Keywords: general relativity; elastic waves; thick shells; speed of propagation; causality; spherical symmetry; Robertson-Walker

\section{Introduction}
\label{Introduction}

Spherically symmetric models with elastic matter in general relativity have been studied by a number of authors: Magli and Kijowski \cite{MK} investigated the problem of elastomechanical equilibrium for a non-rotating star, Park \cite{Park} proved existence theorems for spherically symmetric elastic bodies, Magli \cite{M1} analysed the relativistic interior dynamics of a spherically symmetric non-rotating star composed of an elastic material,
Frauendiener and Kabobel \cite{Fr} discussed spherically symmetric solutions of the general relativistic elasticity equations with different stored energy functions; and Karlovini and Samuelsson \cite{KS1} showed how physically prestressed stellar models, which serve as backgrounds in investigations of stellar perturbations, can be produced numerically and investigated radial and axial perturbations of static spherically symmetric elastic configurations \cite{KS2}, \cite{KS4}; just to name a few.

On the other hand, the study of wave propagation in elastic solids has also been addressed by other authors:  Carter \cite{C} derived a characteristic equation for sound wave fronts in elastic solids using the formalism for the theory of general relativistic elasticity developed in \cite{CQ}, and showed that the characteristic equation can be expressed in terms of the relativistic Hadamard elasticity tensor and can be used to determine the propagation speeds of sound waves in the direction specified by the propagation direction vector and the corresponding polarization directions; Karlovini and Samuelsson \cite{KS1}, who set up a theory of elastic matter sources within the framework of general relativity, also based on the formalism of \cite{CQ}, obtained formulas for the speeds of elastic wave propagation along eigendirections of the pressure tensor and studied these for stiff ultrarigid equations of state \cite{KS3}; Maugin \cite{Maugin} studied wave propagation in prestressed non-linear elastic solids in general relativity; and Kranys \cite{Kranys} determined special relativistic propagation modes for longitudinal and transverse waves of an elastic solid; again, just to name some relevant contributions.

In this paper, after providing a short summary of some relevant
issues concerning relativistic elasticity (the reader is referred to
\cite{BS}, \cite{BCV}, \cite{KS1}, \cite{MK}  for further details),
the definitions of sound wave front, propagation speed and
characteristic equation are presented in Section \ref{sound}. Based
on these definitions, an expression for the propagation speed of the
wave front in spherically symmetric spacetimes with elastic matter
is derived in Section \ref{propagation}, which depends on the energy
density, the radial pressure and the elasticity tensor. In Section
\ref{solutions}, we consider shear-free static and non-static
solutions obtained in \cite {BCV}, determine their radial
propagation speeds and analyse if they satisfy the causality
condition (i.e.: propagation speed less than or equal to the speed
of light), showing that there are open regions where causality is
preserved besides satisfying the Dominant Energy Condition and being
singularity free. Further in Section \ref{matching}, we show that
the non-static solutions referred to above can be matched to
Robertson-Walker spacetimes. We next use two specific non-static
solutions to build up two toy-models: one in which a flat ($k=0$)
Robertson-Walker metric is set as an interior solution and matched
to a spherically symmetric thick shell of elastic material, which in
turn is matched to a non flat ($k=-1$) Robertson-Walker (exterior)
metric; and the other one in which a $k=1$ Robertson Walker interior
is matched to an elastic shell which in turn matched to a flat
Robertson Walker exterior. Continuity of the first and second
fundamental forms on the inner and outer surfaces of the elastic
shell are shown to be satisfied, as well as all the other physically
reasonable conditions: absence of singularities, Dominant Energy
Condition and causal propagation of elastic waves across the elastic
material.

\section{Elasticity in General Relativity. A brief summary}\label{summary}

In order to describe elastic matter in general relativity, one considers a submersion $\psi: M\longrightarrow X,$ from the spacetime manifold $M$ to the three-dimensional material space $X$, which specifies the configuration of the material. The spacetime metric will be denoted by $g$, while $\gamma$ will designate the material metric defined on $X$, which can be thought of as measuring distances between neighbouring particles in the relaxed state of the material. Coordinates in $X$ and $M$ will be denoted by $y^A,$ $A=1,2,3,$ and by $x^a,$ $a=0,1,2,3$, respectively, that is: $y^A = \psi^A(x^b)$.

The differential map $\psi_{\ast}:T_{p}M\longrightarrow T_{\psi(p)}X$ is then surjective and, in the above coordinates, may be represented, at every $p\in M$ by the rank three matrix
\begin{equation}
y^A_b (p)= \left(\frac{\partial y^A}{\partial x^b}\right)_p,\; A=1,2,3;\;\; b=0,1,2,3,
\end{equation}
which is sometimes called \textit{relativistic deformation gradient}. The vector field spanning the kernel of $\psi_{\ast}$, $u^a$, is timelike and can be chosen scaled to unit and future oriented, that is
\begin{equation}
y^{A}_{b}u^{b}=0,\; u^a u_a =-1,\;u^{0}>0,
\end{equation}
and represents the velocity field of the matter in the spacetime $M$.

Note that $\psi$ maps every whole spacetime  trajectory passing through a point $p\in M$ to a single point $\psi(p)$ in the material space $X$ (that is: the world line of a material particle is mapped onto that particle). This fact implies, for instance, that any smooth scalar  field $\phi$ defined on $X$ assigns (via the pull-back $\psi^*$) a real number to each integral line of $u^a$ in $M$, that is: $\Phi= \psi^*\phi$ is a scalar field on $M$ such that $\mathcal{L}_{\vec{u}} \Phi =0$. Further, it defines a one-to-one correspondence between tensor fields $T_{A\ldots}$ defined on the material space $X$ and tensor fields $T_{a\ldots}$ on $M$ satisfying $\mathcal{L}_{\vec{u}} T_{a\ldots}=0$ and $u^a T_{a\ldots}=0$ for any index contraction; that is: all physical magnitudes defined on the material space $X$, can be readily `translated' into spacetime (see, for instance, \cite{Geroch1971} for a detailed proof of this one-to-one correspondence, although in a different context totally unrelated to the present one).

If at a given point $p\in M,$ $k_{ab}\equiv (\psi^{\ast}\gamma)_{ab}=h_{ab},$ where $h_{ab}=g_{ab}+u_a u_b,$  the material is in a locally relaxed state at that point. Otherwise, the material is said to be strained.
The \textit{strain tensor} can be defined as \cite{M2}
\begin{equation}
K_{ab}=k_{ab}-u_a u_b
\end{equation}
and determines the elastic energy stored in an infinitesimal volume element of the material space (energy per particle). That energy will then be a scalar function of $K_{ab}$ and is called \textit{constitutive equation of the material}. The constitutive equation will be represented by $v=v(I_1, I_2, I_3),$ where $I_1 , I_2, I_3$ are the following invariants of $K^{a}_{b}$:
\begin{equation}\label{I1.0} {I_{1}}=\frac{1}{2}\left(\text{Tr} {K}-4\right),\qquad
 {I_{2}}=\frac{1}{4}\left[\text{Tr} {K}^{2}-\left(\text{Tr} {K}\right)^{2}\right]+3,\qquad
 {I_{3}}=\frac{1}{2}\left(\text{det} {K}-1\right).\end{equation}

The energy density $\rho$ is then \be \label{rho0}
\rho = \epsilon v(I_1,I_2,I_3) = \epsilon_0 \sqrt{\det K}\,
v(I_1,I_2,I_3),\ee where $\epsilon_0$ is the particle number density
as measured in the material space, or rather, with respect to the
volume form associated with $k_{ab} = (\psi^* \gamma)_{ab}$, and
$\epsilon$ is that with respect to $h_{ab}$; see \cite{KJ1} for a
proof of the above equation.

The energy-momentum tensor for elastic matter can be expressed as
\begin{equation}
T_{ab}=\rho u_a u_b +p_{ab}=\rho u_a u_b +p h_{ab}+\pi_{ab},
\end{equation}
where $p_{ab}$ represents the \textit{pressure tensor}, $p$ the \textit{isotropic pressure}, and $\pi_{ab}$ the \textit{trace-free anisotropy pressure tensor}. The tensors $p_{ab}$ and $\pi_{ab}$ are symmetric and orthogonal to the flow, $p_{ab}u^a =\pi_{ab}u^a =0$, and $\pi_{ab}$ satisfies also $g^{ab}\pi_{ab} =0$ (trace-free condition).

The energy-momentum tensor can also be written in terms of the invariants of $K^{a}_{b}$ as \cite{M1}
\begin{equation} \label{Tab1}
T^{a}_{\;\;b}=-\rho\,\delta^{a}_{b}+\frac{\partial \rho}{\partial I_{3}}\,
\text{det}K\,h^{a}_{\;\;b}- \left(\text{Tr}
K\,\frac{\partial \rho}{\partial I_{2}}-\frac{\partial \rho}{\partial I_{1}}\right) k^{a}_{\;\;b}+
\frac{\partial \rho}{\partial I_{2}}\,k^{a}_{\;\;c}\,k^{c}_{\;\;b}.
\end{equation}

\section{Speed of sound}\label{sound}

\subsection{Sound wave front and speed of propagation}

Consider an elastic solid with energy-momentum tensor given by
\begin{equation}
T^{ab}=\rho u^a u^b +p^{ab}.
\end{equation}

The conservation law $T^{ab}_{\;\;\; ;b}=0$ implies the following equations of motion
\begin{align}
\rho_{,c}u^c &=-\rho u^{c}_{\; ;c} - p^{cd}u_{c;d}\label{em1}\\
p^{ab}_{\;\;\; ;c}u^c &=2 u^{(a}p^{b)c}\dot{u}_{c}+2p^{c(a}u^{b)}_{\; ;c}-p^{ab}u^{c}_{\; ;c}-E^{abcd}u_{c;d},\label{em2}
\end{align}
where $\dot{u}^a$ is the acceleration vector, namely:
\begin{equation}
\dot{u}^a =u^{a}_{\; ;c}u^c
\end{equation}
and $E^{abcd}$ is the \textit{relativistic elasticity tensor}, which will be defined ahead.

Following \cite{C}, the sound wave fronts are defined as the characteristic hypersurfaces across which the acceleration vector $\dot{u}^a$ can have a jump discontinuity. The map $\psi$ and the metric tensor are assumed to be $C^1$ across these hypersurfaces. Continuity of the first-order derivatives of the map $\psi$ implies that tensors on $M$ arising from the pull-back of tensors on $X$ will be continuous. The velocity $u^a$ is also continuous across the wave fronts.

The normal to the wave front lies in the direction of a vector $\lambda$; thus for instance, $\lambda_a x^a = \mathrm{const.}$ represent plane wave fronts. The vector $\lambda$ can be decomposed as
\begin{equation}
\lambda_a = \nu_a -w u_a.
\end{equation}
where $\nu_a$, the \textit{propagation direction vector}, is a unit, spacelike, transverse vector; that is: $\nu^a\nu_a=1,$ $\nu^a u_a=0.$ In fact, $\lambda = \psi^*(n)$ where $n$ represents the propagation direction 1-form of the wave in the material space.
The scalar $w$ represents the speed of propagation of the wave front with respect to the flow,
\begin{equation}
w=\lambda^a u_a,
\end{equation}
and it must satisfy the local causality condition
\begin{equation}
w^2\leq 1,
\end{equation}
implying that the characteristic hypersurface must be timelike or null (units are taken so that the speed of light is $c=1$); or else $\lambda_a\lambda^{a} \geq 0$ (spacelike or null).

The acceleration discontinuity can be expressed as
\begin{equation}\label{ad}
[\dot{u}^a]=\alpha\iota^a,
\end{equation}
where $\alpha$ is the amplitude of the wave front and $\iota^a$ is the \textit{polarization vector} of the wave front, satisfying $\iota^a\iota_a=1$ and $\iota^a u_a=0,$ since $\dot{u}^a u_a =0.$


\subsection{Characteristic equation}

The \textit{relativistic elasticity} tensor $E^{abcd}$ is a bisymmetric tensor function of state \cite{KS1} defined as
\begin{equation}
E^{abcd}=-2\frac{\partial p^{ab}}{\partial g_{cd}} - p^{ab}h^{cd}.
\end{equation}
It satisfies the symmetry conditions
\begin{equation}
E^{abcd}=E^{(ab)(cd)}=E^{cdab}
\end{equation}
and is orthogonal to the velocity of the flow,
\begin{equation}
E^{abcd}u_{d}=0.
\end{equation}
The elasticity tensor can be rewritten as
\begin{equation}
E^{abcd}=-2\epsilon \frac{\partial}{\partial h_{cd}}\left(\frac{p^{ab}}{\epsilon}\right),
\end{equation}
or, equivalently,
\begin{equation}\label{ET3}
E^{abcd}=4\epsilon \frac{\partial^2 v}{\partial h_{ab} h_{cd}}=4\epsilon \frac{\partial^2 v}{\partial g_{ab} g_{cd}},
\end{equation}
where, as defined previously, $\epsilon$ stands for the particle number density and $v$ represents the constitutive function.

The relativistic Hadamard elasticity tensor is defined in terms of the elasticity tensor by
\begin{equation}\label{A}
A^{abcd}=E^{abcd}-h^{ac}p^{bd}.
\end{equation}
This tensor has the symmetry
\begin{equation}
A^{abcd}=A^{cdab}
\end{equation}
and is also orthogonal to the velocity of the flow
\begin{equation}
A^{abcd}u_d=A^{abdc}u_d=0.
\end{equation}
From \eqref{em1} and \eqref{em2}, Carter \cite{C} derived the following characteristic equation
\begin{equation}
\{w^2 (\rho h^{ac}+p^{ac})-Q^{ac}\}\iota_c =0,\label{ce}
\end{equation}
which depends on the so called  \textit{relativistic Fresnel tensor} $Q^{ac}$, defined as
\begin{equation}\label{Q}
Q^{ac}=A^{abcd}v_bv_d,
\end{equation}
which is symmetric,
\begin{equation}
Q^{ac}=Q^{(ac)}
\end{equation}
and flowline orthogonal
\begin{equation}
Q^{ac}u_{c}=0.
\end{equation}

\section{Applications to spherically symmetric elastic spacetimes}\label{propagation}

\subsection{Spacetime configuration and elasticity tensor}

Consider a spherically symmetric spacetime  $(M,g)$, $M$ being a 4-dimensional Hausdorff, simply connected manifold of class $\mathcal{C}^2,$ with metric $g$ given by the following line-element
\begin{equation}
\label{m1}
ds^2=-a^2 dt^2+b^2 dr^2+Y^2(d\theta^2+\sin^2 \theta d\phi^2),
\end{equation}
where $a$, $b$ and $Y$ are functions of the coordinates $t$ and $r$. 

It can be shown (see \cite{BCV}) that the above coordinates and form of the metric can be chosen so that the velocity flow is comoving with the time coordinate; thus:
defining the tetrad $\{u,e_1,e_2,e_3\}$, where $$u^a=\left(a^{-1},0,0,0\right)$$ is the velocity vector of the flow and $$e_{1}^{a}=\left(0,b^{-1},0,0\right),\;e_{2}^{a}=(0,0,Y^{-1},0),\;e_{3}^{a}=(0,0,0,(Y\sin \theta)^{-1}),$$ the metric can be written as
$g_{ab}=-u_a u_b +e_{1a} e_{1b} +e_{2a}e_{2b} +e_{3a}e_{3b}.$

In \cite{BCV} the authors considered solutions for the case where the line element of the material metric $\gamma$ corresponding to
\eqref{m1} is
\begin{equation}
\label{m2}
d\sigma^2 = f^2(r)\left[dr^2+r^2(d\theta^2 +\sin^2 \theta d\phi^2)\right],
\end{equation}
that is: the most general form for a 3-dimensional, spherically symmetric metric.

The pulled-back material metric, $k_{ab}$, is such that $k^{a}_{b}$ has two different eigenvalues:
\begin{equation}
\label{ev} s=
f^{2}(r)\,\frac{r^{2}}{Y^{2}},\qquad \eta=\frac{f^{2}(r)}{b^2},
\end{equation} $s$ having algebraic multiplicity two.
The invariants \eqref{I1.0} can then be written in terms of those eigenvalues as
\begin{equation}
\label{I1}I_{1}=
\frac{1}{2}\left(\eta + 2 s -3\right),\qquad
I_{2}=
 -\frac12 \left( s^2 +2 \eta s +\eta +2 s\right) -3,\qquad
I_{3}=
\frac12\left( \eta s^2 -1\right),
\end{equation}
and the rest frame energy per unit volume takes the form
\be\label{rho}
\rho=\epsilon v=\,\epsilon_{0}\,s\,\sqrt{\eta}\,v(s,\eta),
\ee
$v=v(s,\eta)$ being the constitutive equation.
The energy-momentum tensor \eqref{Tab1} has the following non-zero components
\begin{equation}\begin{split}
\label{T2}
&T^{0}_{0}= -\epsilon v,\\
&T^{1}_{1}=2\,\epsilon\,\eta\,\frac{\partial v}{\partial\eta},\\
&T^{2}_{2}=\epsilon\,s\,\frac{\partial v}{\partial s}.
\end{split}\end{equation}

In this context, static and non-static shearfree spherically symmetric solutions were presented, along with the corresponding field equations $G^{a}_{b}=8\pi T^{a}_{b}$, and were shown to satisfy the Dominant Energy Condition (DEC) in open regions of the spacetime. We will consider these solutions in section \ref{solutions} of the present paper, analysing whether or not the propagation of elastic waves has a causal behaviour within those spacetime regions where the DEC holds.

Before doing so, we shall introduce some auxiliary results which will be of use to that end.

First, we shall obtain a useful expression for the relativistic elasticity tensor $\displaystyle{E^{abcd}=4\epsilon \frac{\partial^2 v}{\partial g_{ab} g_{cd}}}$.

Using
\begin{equation}\label{dg}
\frac{\partial}{\partial g_{ab}}=-g^{ac}g^{bd}\frac{\partial}{\partial g^{cd}},
\end{equation}
and also (see, e.g. \cite{KS1}):
\begin{equation}\label{gk}
\frac{\partial}{\partial g^{cd}}=\frac{1}{2}\left(k_{mc}\frac{\partial }{\partial k^{d}_{m}}+k_{md}\frac{\partial }{\partial k^{c}_{m}}\right),
\end{equation}
one obtains:

\begin{equation}\label{vg}
\frac{\partial v}{\partial g_{ab}}=-\frac{1}{2}\left[g^{bd}k_{m}^{a}\frac{\partial v}{\partial k^{d}_{m}}+g^{ac}k_{m}^{b}\frac{\partial v}{\partial k^{c}_{m}}\right].
\end{equation}
Now, for the spherically symmetric case, we can write $k^{a}_{b}$ in terms of the tetrad vectors as
\begin{equation}
k^{a}_{b}=\eta e_{1}^{a}e_{1b}+se_{2}^{a}e_{2b}+se_{3}^{a}e_{3b},
\end{equation}
whence the eigenvalues can be extracted as
$\eta=k^{a}_{b}e_{1a}e^{1b}$ and $s=\frac{1}{2}k^{a}_{b}(e_{2a}e_{2}^{b}+e_{3a}e_{3}^{b})$.
Then one calculates
\begin{align}
\frac{\partial v}{\partial k^{d}_{m}}=\frac{\partial v}{\partial \eta}\frac{\partial \eta}{\partial k^{d}_{m}}+\frac{\partial v}{\partial s}\frac{\partial s}{\partial k^{d}_{m}}
=\frac{\partial v}{\partial \eta}e_{1d}e_{1}^{m}+\frac{1}{2}\frac{\partial v}{\partial s}(e_{2d}e_{2}^{m}+e_{3d}e_{3}^{m}).
\end{align}
Substituting this expression into \eqref{vg} gives
\begin{equation}\label{dvg}
\frac{\partial v}{\partial g_{ab}}=-\eta\frac{\partial v}{\partial \eta}e_{1}^{a}e_{1}^{b}-\frac{1}{2}s\frac{\partial v}{\partial s}(e_{2}^{a}e_{2}^{b}+e_{3}^{a}e_{3}^{b}).
\end{equation}
Now, one only needs to calculate $\frac{\partial}{\partial g_{cd}}\left(\frac{\partial v}{\partial g_{ab}}\right)$, cf. \eqref{ET3}.
For that purpose, we derive the following expressions using \eqref{gk}
\begin{equation}\label{ddveta}
\frac{\partial}{\partial g_{cd}}\left(\frac{\partial v}{\partial \eta}\right)=-\eta\frac{\partial^2 v}{\partial \eta^2}e_{1}^{c}e_{1}^{d}-\frac{1}{2}s\frac{\partial^2 v}{\partial \eta \partial s}(e_{2}^{c}e_{2}^{d}+e_{3}^{c}e_{3}^{d}),
\end{equation}
\begin{equation}\label{ddvs}
\frac{\partial}{\partial g_{cd}}\left(\frac{\partial v}{\partial s}\right)=-\eta\frac{\partial^2 v}{\partial s \partial \eta}e_{1}^{c}e_{1}^{d}-\frac{1}{2}s\frac{\partial^2 v}{\partial s^2}(e_{2}^{c}e_{2}^{d}+e_{3}^{c}e_{3}^{d}),
\end{equation}
\begin{equation}\label{deta}
\frac{\partial \eta}{\partial g_{cd}}=-\eta e_{1}^{c}e_{1}^{d},
\end{equation}
\begin{equation}\label{ds}
\frac{\partial s}{\partial g_{cd}}=-\frac{1}{2}s (e_{2}^{c}e_{2}^{d}+e_{3}^{c}e_{3}^{d}).
\end{equation}
Taking into account the following result, presented in \cite{KS1},
\begin{equation}
\frac{\partial e_{\rho}^{c}}{\partial g^{ab}}=\frac{1}{2}e_{\rho}^{c}e_{\rho a}e_{\rho b}+\sum_{\sigma\neq \rho}\frac{n^{2}_{\rho}}{n^{2}_{\rho}-n^{2}_{\sigma}}e_{\sigma}^{c}e_{\rho (a}e_{\sigma b)},
\end{equation}
where $n_{\sigma}^{2}$ and $n_{\rho}^{2}$, $\sigma,\rho=1,2,3,$ are the eigenvalues of $k^{a}_{b}$ (in our case $n_{1}^{2}=\eta$, $n_{2}^{2}=n_{3}^{2}=s$), together with \eqref{dg},
yields
\begin{equation}\label{e1}
\frac{\partial }{\partial g_{cd}}\left(e_{1}^{a}e_{1}^{b}\right)=-e_{1}^{a}e_{1}^{b}e_{1}^{c}e_{1}^{d}-\frac{2\eta}{\eta-s}
\left(e_{1}^{(a}e_{2}^{b)}e_{1}^{(c}e_{2}^{d)}+e_{1}^{(a}e_{3}^{b)}e_{1}^{(c}e_{3}^{d)}\right),
\end{equation}
\begin{equation}\label{e2}
\frac{\partial }{\partial g_{cd}}\left(e_{2}^{a}e_{2}^{b}\right)=-e_{2}^{a}e_{2}^{b}e_{2}^{c}e_{2}^{d}-\frac{2s}{s-\eta}
e_{1}^{(a}e_{2}^{b)}e_{2}^{(c}e_{1}^{d)},
\end{equation}
\begin{equation}\label{e3}
\frac{\partial }{\partial g_{cd}}\left(e_{3}^{a}e_{3}^{b}\right)=-e_{3}^{a}e_{3}^{b}e_{3}^{c}e_{3}^{d}-\frac{2s}{s-\eta}
e_{1}^{(a}e_{3}^{b)}e_{3}^{(c}e_{1}^{d)}.
\end{equation}
Finally, from \eqref{dvg}, and using \eqref{ddveta}-\eqref{ds} and \eqref{e1}-\eqref{e3}, one finally obtains the following expression for the elasticity tensor
\begin{align}\label{et}
E^{abcd}=4\epsilon\frac{\partial^2 v}{\partial g_{ab}\partial g_{cd}}
&=4\epsilon\left[\left(2\eta \frac{\partial v}{\partial \eta}+\eta^2 \frac{\partial^2 v}{\partial \eta^2}\right)\frac{1}{b^4}\delta_{r}^{a}\delta_{r}^{b}\delta_{r}^{c}\delta_{r}^{d}\right.\nonumber\\
&+\left(\frac{2\eta^2}{\eta-s}\frac{\partial v}{\partial \eta}+\frac{s^2}{s-\eta}\frac{\partial v}{\partial s}\right)\left(\frac{1}{b^2Y^2}\delta_{r}^{(a}\delta_{\theta}^{b)}\delta_{r}^{(c}\delta_{\theta}^{d)}+
\frac{1}{b^2Y^2 \sin^2\theta}\delta_{r}^{(a}\delta_{\phi}^{b)}\delta_{r}^{(c}\delta_{\phi}^{d)}\right)\nonumber\\
&+ \frac{1}{2}\eta s \frac{\partial^2 v}{\partial \eta \partial s}
\left(\frac{1}{b^2Y^2}(\delta_{r}^{a}\delta_{r}^{b}\delta_{\theta}^{c}\delta_{\theta}^{d}+\delta_{r}^{c}\delta_{r}^{d}\delta_{\theta}^{a}\delta_{\theta}^{b})+
\frac{1}{b^2Y^2 \sin^2\theta}(\delta_{r}^{a}\delta_{r}^{b}\delta_{\phi}^{c}\delta_{\phi}^{d}+\delta_{r}^{c}\delta_{r}^{d}\delta_{\phi}^{a}\delta_{\phi}^{b})\right)\nonumber\\
&+\left(\frac{3}{4}s\frac{\partial v}{\partial s}+\frac{1}{4}s^2\frac{\partial^2 v}{\partial s^2}\right)\left(\frac{1}{Y^4}\delta_{\theta}^{a}\delta_{\theta}^{b}\delta_{\theta}^{c}\delta_{\theta}^{d}+
\frac{1}{Y^4 \sin^4\theta}\delta_{\phi}^{a}\delta_{\phi}^{b}\delta_{\phi}^{c}\delta_{\phi}^{d}\right)\nonumber\\
&\left.+\left(\frac{1}{4}s\frac{\partial v}{\partial s}+\frac{1}{4}s^2\frac{\partial^2 v}{\partial s^2}\right)\frac{1}{Y^4\sin^2 \theta }\left(\delta_{\theta}^{a}\delta_{\theta}^{b}\delta_{\phi}^{c}\delta_{\phi}^{d}+
\delta_{\phi}^{a}\delta_{\phi}^{b}\delta_{\theta}^{c}\delta_{\theta}^{d}\right)\right].
\end{align}

\subsection{Second-order and fourth-order tensors}
Consider the set of all symmetric, second order tensors, that are spherically symmetric and orthogonal to the flow vector $u^a$.
Let $S_{ab}$ be one such tensor field, then it satisfies:
\begin{enumerate}
\item $S_{ab}=S_{ba};$
\item $S_{ab}u^{b}=0;$
\item $\mathcal{L}_{\vec{\xi}_{A}}S_{ab}=0,$ where $\vec{\xi}_A$, $A=1,2,3,$ are the usual Killing vectors implementing the spherical symmetry; namely
$$\vec{\xi}_1 = \cos\phi\,\partial_\theta-\sin\phi\cot\theta\,\partial_\phi, \;\; \vec{\xi}_2 = -\sin\phi\,\partial_\theta-\cos\phi\cot\theta\,\partial_\phi, \;\; \vec{\xi}_3 = \partial_\phi.$$
\end{enumerate}
One can then write
\begin{equation}
S_{ab}dx^a dx^b =\alpha(t,r)dr^2 + \beta(t,r)(d\theta^2 + \sin^2 \theta d\phi^2).
\end{equation}
It is immediate to see that the set of all tensor fields with the above properties, defines at each spacetime point a 2 dimensional vector space, a basis for which is  $\{h_{ab},p_{ab}\}$ at that point, with
\begin{equation}
h_{ab}dx^a dx^b =b^2dr^2+Y^2(d\theta^2 + \sin^2 \theta d\phi^2),\;\; p_{ab}dx^a dx^b =P_1 dr^2+P_2(d\theta^2 + \sin^2 \theta d\phi^2),
\end{equation}
where the case in which the pressure tensor $p_{ab}$ is proportional to $h_{ab}$ is explicitly ruled out, as it would correspond to a perfect fluid.

Thus, a tensor field $S_{ab}$ with the above properties may be written as
\begin{equation}\label{Sab}
S_{ab}=A h_{ab}+B p_{ab},
\end{equation}
where $A=A(t,r)$ and $B=B(t,r)$.


Consider now a fourth order tensor $E^{abcd}$ having the following properties
\begin{enumerate}
\item $E^{abcd}=E^{cdab};$
\item $E^{abcd}=E^{bacd}=E^{abdc};$
\item $E^{abcd}u_{d}=0;$
\item $\mathcal{L}_{\vec{\xi}_{A}}E^{abcd}=0$, where, as before, $\vec{\xi}_A$, $A=1,2,3,$ designate the Killing vectors implementing spherical symmetry.
\end{enumerate}
It is then easy to see that, with the conventions set up above, $E^{abcd}$ can be written as
\begin{equation}
E_{abcd}=E_1 h_{ab}h_{cd}+E_2 (h_{ab}p_{cd}+p_{ab}h_{cd})+E_3 p_{ab} p_{cd},
\end{equation}
where $E_1, E_2, E_3$ are functions of $t$ and $r$.

\subsection{The speed of propagation}
In spherical symmetry, since $\dot{u}^a \propto \partial_{r},$ it follows from \eqref{ad} that
$\nu^a=\iota^a.$

Further, the propagation direction vector $\nu^a$ satisfies
\begin{equation}\label{nu1}
\nu^a\propto \partial_{r} \;\; \text{and therefore}\;\;\nu^a=\left(0,b^{-1},0,0\right).
\end{equation}
This follows follows from the requirement of spherical symmetry, which imposes $\mathcal{L}_{\vec{\xi}_A} u^a = 0$ and  $\mathcal{L}_{\vec{\xi}_A} \lambda_a = 0$. Since $\nu_a u^a = 0$, one has that $u^a\mathcal{L}_{\vec{\xi}_A} \nu_a =0$, and then also: $\mathcal{L}_{\vec{\xi}_A} \nu_a - u_a \mathcal{L}_{\vec{\xi}_A} w =0$, contracting this last equality with $u^a$, one gets $\mathcal{L}_{\vec{\xi}_A} w =0$, and then $\mathcal{L}_{\vec{\xi}_A}\nu_a = 0$, which implies \eqref{nu1}.

Since the relativistic Fresnel tensor has the following properties
\begin{equation}
Q^{ab}=Q^{(ab)},\;\; Q^{ab}u_b =0,\;\; \mathcal{L}_{\vec{\xi}_A} Q^{ab}=0,
\end{equation}
one can write it in the form (see \eqref{Sab}):
\begin{equation}
Q_{ab}=\alpha h_{ab}+\beta p_{ab}.
\end{equation}

Consequently, the characteristic equation \eqref{ce} can be expressed as
\begin{equation}
\left\{w^2 (\rho h^{ac}+p^{ac})-(\alpha h^{ac}+\beta p^{ac})\right\}\iota_c =0
\end{equation}
and, using $\nu_{c}=\iota_{c}= b \delta_{c}^{r},$
one obtains
\begin{align}
w^2 (\rho h^{ar} b+p^{ar}b)-(\alpha h^{ar} b+\beta p^{ar} b)=0,\; \text{ for } a\neq r,\label{w21}\\
w^2 \left(\frac{\rho}{b}+b p^{rr}\right)-\left(\frac{\alpha}{b}+\beta b p^{rr}\right) =0,\; \text{ for } a= r.\label{w22}
\end{align}
Equation \eqref{w21} is trivially satisfied since both $h^{ab}$ and $p^{ab}$ are diagonal.
From \eqref{w22} one concludes
\begin{equation}
w^2=\frac{b Q^{rr}}{\frac{\rho}{b}+b p^{rr}},
\end{equation}
which, using \eqref{A} and \eqref{Q} can be written as
\begin{equation}
w^2=\frac{b^3E^{rrrr}-b p^{rr}}{\frac{\rho}{b}+b p^{rr}}.
\end{equation}
From \eqref{et} one gets
\begin{equation}
E^{rrrr}=\frac{4\epsilon}{b^4}\left(2\eta\frac{\partial v}{\partial\eta}+\eta^2 \frac{\partial^2 v}{\partial \eta^2}\right)
\end{equation}
and
\begin{equation}
p^{rr}=T^{rr}=2 \epsilon \eta\frac{\partial v}{\partial\eta}\frac{1}{b^2},
\end{equation}
so that in the present case the speed of propagation of the wave front is
\begin{align}\label{w2}
w^2&=
\frac{6\epsilon \eta\frac{\partial v}{\partial\eta}+ 4 \epsilon\eta^2 \frac{\partial^2 v}{\partial \eta^2}}{\epsilon v +2 \epsilon\eta \frac{\partial v}{\partial \eta}}=\frac{3T^{1}_{1}+ 4 \epsilon\eta^2 \frac{\partial^2 v}{\partial \eta^2}}{-T^{0}_{0}+T^{1}_{1}}.
\end{align}

It is worth noting that \eqref{w2} obtained above is in agreement with the expression given in \cite{KS1} for the propagation velocity of the elastic waves.

\section{Specific solutions} \label{solutions}

We will now explore whether the condition $0 \leq w^2 \leq 1$ is satisfied for the static and non-static shearfree solutions presented in \cite{BCV}.
While we do not claim that the solutions presented here have any particular relevance, we would like to point out to the fact that it is possible to find exact solutions, with elastic material content, such that are singularity free, satisfy the DEC, and behave causally when perturbed (i.e.: the speed of sound of the elastic waves is less than the speed of light), as the following examples show.

\subsection{Shear-free static solution}

Consider the static shear-free solution obtained in \cite{BCV}, which is a subcase of \eqref{m1} with
$a^2=\frac{1}{Y^2},$ $b^2=Y^2,$ where $Y=e^{-\frac{5}{2}r^2}.$
In this case, the energy density $\rho$, the radial pressure $p_{1}$ and the tangential pressure $p_2$ are, respectively,
\begin{align}
\rho &= - T^{0}_{0}= \epsilon v = \frac1{8\pi} e^{5r^2}(11-25 r^2),\\
p_1 &=T^{1}_{1}=2\epsilon \eta\frac{\dep v}{\dep\eta}= -
\frac1{8\pi}e^{5r^2}(25 r^2 + 1),\label{p1static}\\
p_2 &=T^{2}_{2}= \epsilon s\frac{\dep
v}{\dep s} =\frac1{8\pi} 25r^2 e^{5r^2}.\label{example1.3}
\end{align}
This solution satisfies the dominant energy condition
for $r \in \left[0, \frac{1}{\sqrt{5}}\right)$
and is non-singular at the origin.

Calculating $w^2$ from \eqref{w2}, using \eqref{ev} with $f(r)=\frac{e^{-\frac{5}{2}r^2}}{(75r^2+1)^{\frac{1}{3}}}$
and \eqref{rho}, and the expression for $\frac{\partial^2 v}{\partial^2 \eta}$ obtained from \eqref{p1static}we get:
\begin{align}
\frac{\partial^2 v}{\partial^2 \eta}=\frac{e^{5r^2}(75r^2 +1)^{\frac{7}{3}}}{800\pi \epsilon_0 r^4}\left(9375 r^6 +3625 r^4 +55 r^2 -1\right),
\end{align}
that gives
\begin{equation}
w^2=\frac{9375r^6+1750r^4-20r^2-1}{-1250r^4+250r^2}.
\end{equation}

\begin{figure}[H]
\begin{center}
\includegraphics[width=8cm]{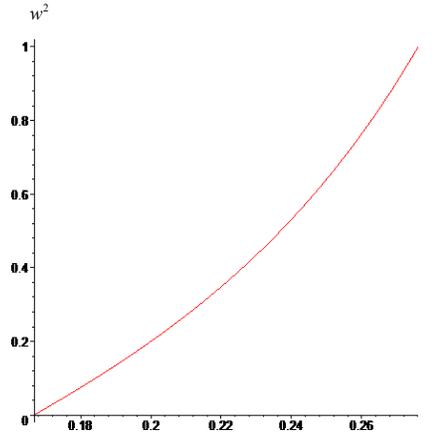}\\
\caption{Graph of $w^2$ for $r \in \left(0.167,0.276\right)$.}
\label{f1}
\end{center}
\end{figure}

The condition $0 \leq w^2 \leq 1$ is satisfied for $r\in
(a,b)\subset \left(0, \frac{1}{\sqrt{5}}\right),$ where $a\approx
0.167$ and $b\approx 0.276$ (see Figure \ref{f1}).

We note that in \cite{BCV} it was shown that this solution can be
used to build up a static two-layer star model consisting of an
elastic core, surrounded by a perfect fluid corresponding to the
interior Schwarzschild solution matched to the vacuum Schwarzschild
solution.

\subsection{Non-static, shear-free  solutions}
\label{secnssol}

In order to analyse the condition $0 \leq w^2 \leq 1$ for the non-static shearfree solutions presented in \cite{BCV}, whose spacetime  metrics are of the form
\begin{equation}
\label{shearfree3}
ds^2 = - dt^2 +t^2 B^2(r)\left(dr^2 + d \theta^2 + \sin^2 \theta d\phi^2 \right),
\end{equation}
we will first derive an expression for $w^2$ in terms of the coordinates $t,r$ and the functions $B$ and its first and second derivatives with respect to $r$ (noted as $B'$ and $B''$ respectively). Notice that this metric can be obtained from \eqref{m1} replacing the metric functions by $a^2 =1$ and $b^2=Y^2=t^2 B^2$.

In this case, the energy density and the radial and tangential pressures take the form
\begin{align}
\rho &=-T^{0}_{0}=\frac1{8\pi\, t^2}\left(- \frac{2B''}{B^3} + \frac{B'^2}{B^4} + \frac{1}{B^2} + 3  \right),\label{rho1}\\
p_1 &=T^{1}_{1}= \frac1{8\pi\, t^2} \left( \frac{B'^2}{B^4} - \frac{1}{B^2} - 1  \right),\label{p1}\\
p_2 &=T^{2}_{2}= \frac1{8\pi\, t^2} \left( \frac{B''}{B^3}- \frac{B'^2}{B^4} -  1  \right).
\end{align}

The term $4 \epsilon\eta^2 (\partial^2 v/\partial \eta^2)$ in \eqref{w2} can be calculated using \eqref{rho} and \eqref{ev}, with $f(r)$ satisfying (see \cite{BCV}) $$\frac{f'}{f} = \frac{B'}{B^2}-\frac{2}{3rB},$$ and applying the inverse function theorem, which yields:
\begin{align}
\frac{\partial t}{\partial \eta}&=-\frac{t^3}{2f^2}\left(B^2-\frac{2}{3}B+rB'-rB'B\right)   ,\\
\frac{\partial r}{\partial \eta}&= -\frac{rt^2B^2}{2f^2},
\end{align}
the result being then
\begin{align}
4 \epsilon\eta^2 \frac{\partial^2 v}{\partial \eta^2}= &\frac{1}{4\pi t^2 B^2}\left[\left(\frac{1}{2}+\frac{2}{3B}\right)\left(-\frac{B'^2}{B^2}+B^2+1\right)\right.\nonumber\\ & +\left.\frac{rB'}{B^2}   \left(-B''+\frac{B'^2}{B^2}+\frac{B'^2}{B}+B^3 -B^2 -1\right)\right].\label{dv2}
\end{align}
Substituting \eqref{rho1}, \eqref{p1} and \eqref{dv2} in \eqref{w2} yields
\begin{align}
w^2=\frac{\left(\frac{2}{3}-B\right)\left(-\frac{B'^2}{B}+B^3+B\right)+ rB'   \left(-B''+\frac{B'^2}{B^2}+\frac{B'^2}{B}+B^3 -B^2 -1\right)}{-B B''+B'^2+B^4}.
\end{align}

We next present two specific examples.

{\bf Example 1}

\begin{figure}[H]
\begin{center}
\includegraphics[width=8cm]{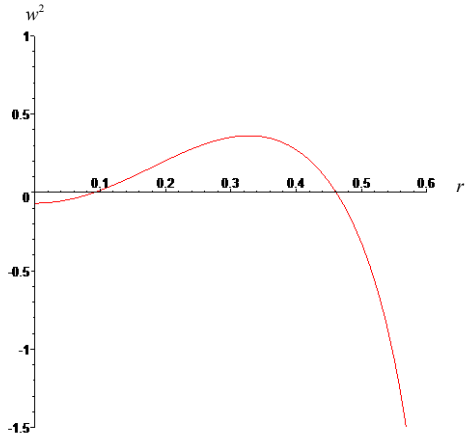}\\
\caption{ Graph of $w^2$ for $c=1$, $r_{0}=0$, where $0\leq w^2
\leq1$ for  $r \in \left(0.092,0.461\right)$.} \label{f2}
\end{center}
\end{figure}

Consider the solution obtained by substituting
\begin{equation}\label{example1}
B(r) = \frac{\sqrt{3}}{9c} \left[-1 + 3 \cosh^2 \left(\frac{r-r_0}{c}\right)\right]^{\frac32}
\end{equation}
into \eqref{shearfree3}, $c\neq 0$ and $r_0$ being real constants.
The DEC is satisfied for certain ranges of the radial coordinate $r\in [0, R)$.
Specifying $c$ and $r_0,$ one can find an interval for $r,$ where $0\leq w^2 \leq 1$ is valid.
The example presented in Figure \ref{f2} illustrates this fact.

{\bf Example 2}

\begin{figure}[H]
\begin{center}
\includegraphics[width=8cm]{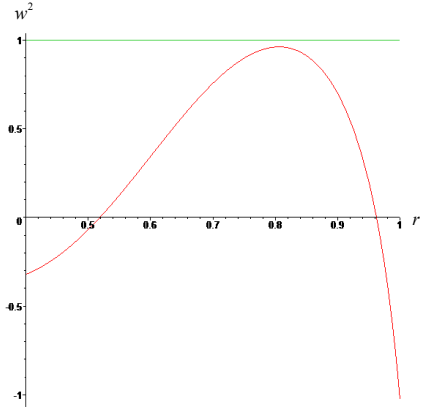}\\
\caption{ \textsl{Graph of $w^2$ for $r_{0}=1/2$, where $0\leq w^2
\leq1$ for $r \in \left(0.519,0.962\right)$.}} \label{f3}
\end{center}
\end{figure}

In this example, the solution is given by the metric \eqref{m1} with
\begin{equation}\label{example2}
B(r) = \frac{\sqrt{3}}{9} \left(2 + 3 (r-r_0)^2\right)^\frac32,
\end{equation}
where $r_0 \neq 0$ is a real constant. The DEC is satisfied for
certain ranges of the radial coordinate $r\in [0, R)$. Choosing a
value for $r_0$, one can find intervals for $r$, where $0\leq w^2
\leq 1$ is satisfied (see Figure \ref{f3}).

\section{Matching of non-static elastic and Robertson-Walker spacetimes}\label{matching}

We next show that the non-static shearfree elastic solutions (\ref{shearfree3}) can be matched to
certain Robertson-Walker (RW) spacetimes.

The Robertson-Walker line element in its isotropic form is given by
(see for instance \cite{Fay},\cite{Fay1})
\begin{equation}
ds^{2} = - dT^2 +\frac{a^2(T)}{b^2(R)}\left ( dR^2 + R^2 d\Omega^2\right),\label{RW1}
\end{equation}
where $b(R)=1+\frac{k}{4}R^2$ with $k=0,\pm 1,$
and the mass function and Friedmann equations are
\begin{equation}
m(T,R)=\frac{\rho}{6}\left(\frac{Ra}{b}\right)^3,
\end{equation}
\begin{align}
\rho=\frac{3(\dot{a}^2+k)}{a^2},\;\;\;\dot{\rho}+3(\rho+p)\frac{\dot{a}}{a}=0,
\end{align}
where a dot indicates derivative with respect to the time $T$.

\subsection{Case: RW with $k=0$ and $a(T)=\alpha T,$ $\alpha>0$}
\label{section1}

Let us now consider the junction across a spherically symmetric surface $\Sigma$ of the elastic, non-static solution
\begin{equation}\label{est}
ds_{-}^{2} = - dt^2 + t^2B^2(r)(dr^2+ d\Omega^2),
\end{equation}
with the Robertson-Walker metric
\begin{equation}
ds_{+}^{2} = - dT^2 +\alpha^2 T^2\left( dR^2 + R^2 d\Omega^2\right),\label{RW01}
\end{equation}
which is considered in \cite{Fay}.

The signs $-$ and $+$ are used to denote, respectively, interior and exterior quantities. At this point, it should be emphasised though, that the labels \lq interior' and \lq exterior' are quite arbitrary, nevertheless, we shall use them just as a convenient way of referring to the spacetimes considered.

For the metric \eqref{RW01} one has
\begin{equation}\label{RWrhop01}
\rho=\frac{3}{T^2},\;\; p=-\frac{1}{T^2}
\end{equation}
and
\begin{equation}\label{RWm01}
m(T,R)=\frac{\alpha^3 TR^3}{2}.
\end{equation}
Notice that this is a particular case of a flat Robertson-Walker metric with linear equation of state $p=\gamma \rho$, where $\gamma = -\frac{1}{3},$ which satisfies the Dominant Energy Condition (see \cite{Fay}).

The boundary $\Sigma^-$ can be parametrized by $\{ t = \lambda, r = r_\Sigma \}$, where
$r_{\Sigma}$ is a constant. The tangent space to $\Sigma^{-}$ at any of its points, is spanned by the following orthogonal tangent vector fields at that point
\begin{equation}
T\Sigma^{-}=\langle e_{1}^{-}=\partial_t,e_{2}^{-}=\partial_{\theta},e_{3}^{-}=\partial_{\phi}\rangle.
\end{equation}
The outgoing, unit, normal vector field to $\Sigma^{-}$ is given by
\begin{equation}
n^{a-}=t^{-1}B^{-1}\partial_{r}.
\end{equation}

Consider next the boundary $\Sigma^+$; it can be parametrized by $\{ T = \lambda, R = R_\Sigma \}$,
where $R_\Sigma $ is a constant.
The tangent space to $\Sigma^{+}$ can be generated by the following orthogonal tangent vectors
\begin{equation}
T\Sigma^{+}=\langle e_{1}^{+}=\partial_T
,e_{2}^{+}=\partial_{\theta},e_{3}^{+}=\partial_{\phi}\rangle, \label{tangentspaceRW}
\end{equation}
and the unit normal vector field to $\Sigma^{+}$ is
\begin{equation}
n^{a+}=\frac{1}{\alpha T}\partial_{R}.
\end{equation}

Now, the two spacetimes can be matched across $\Sigma$ if and only the first and second fundamental forms of their respective metrics are continuous across it (see for instance \cite{Fay}).

The first fundamental forms on $\Sigma$ are given by
$$q_{\alpha\beta}^{\pm}=e_{\alpha}^{\pm a}e_{\beta}^{\pm b}g_{ab}^{\pm},\;\;\alpha,\beta=1,2,3,$$ and a trivial calculation yields for their corresponding line elements
\begin{align}
d\sigma_{-}^2& =-d\lambda^2+t^{2} B^2(r_\Sigma) d\Omega^2,\label{FF1010}\\
d\sigma_{+}^{2}& =-d\lambda^2+\alpha^2\lambda^2 R_\Sigma^2 d\Omega^2,
\end{align}
where all quantities must be evaluated on $\Sigma$.

The continuity of the first fundamental form (first matching condition) $q_{\alpha\beta}^{-}=q_{\alpha\beta}^{+}$, implies then
\begin{align}
\lambda^2 B^2 &\stackrel{\Sigma}{=}\alpha^2 \lambda^2 R_\Sigma^2,\label{FF101}
\end{align}
where $\stackrel{\Sigma}{=}$ means that the equality holds only at points on the surface $\Sigma$, (recall that, on $\Sigma$, one has $t=T=\lambda$).

The independent components of the second fundamental forms,
$$H^{\pm}_{\alpha\beta}=-n_{a}^{\pm}e_{\alpha}^{\pm b}\nabla_{b}^{\pm}e_{\beta}^{\pm a},$$
at points on $\Sigma$ are (cf. \cite{CT})
\begin{align}
&H_{11}^{-}=0,\;\;\;H_{22}^{-}=\lambda B', \label{FF2010}\\
&H_{11}^{+}=0,\;\;\;H_{22}^{+}=\alpha \lambda R_\Sigma,
\end{align}
where again, all quantities are evaluated on $\Sigma$.
Continuity of the second fundamental forms (second matching condition), $H^{+}_{\alpha\beta}=H^{-}_{\alpha\beta}$, implies
\begin{align}
B'\stackrel{\Sigma}{=} \alpha  R_\Sigma.\label{FF201}
\end{align}

Thus, from \eqref{FF101} and \eqref{FF201}, it follows that the elastic spacetime \eqref{est} can be matched to a flat Robertson-Walker spacetime of the form \eqref{RW01} if and only if
\begin{align}
B\stackrel{\Sigma}{=} \alpha R_\Sigma,\;\;\;B'\stackrel{\Sigma}{=}\alpha R_\Sigma\label{FF01}.
\end{align}

Notice that, if these conditions are satisfied, it readily follows from \eqref{p1} (using units such that $8\pi G=c=1$) that
\begin{equation}
p_1^{-} = -\frac{1}{t^2}\;\; \text{and then}\;\; p_1^{-}\stackrel{\Sigma}{=}-\frac{1}{\lambda^2},
\end{equation}
and one then has (cf. (\ref{RWrhop01}))
\begin{equation}
p_1^{-} \stackrel{\Sigma}{=}p_1^{+},
\end{equation}
which is a well-known necessary (but not sufficient) condition for the matching of two spacetimes, namely, pressure orthogonal to the matching surface must be continuous across it (which follows from the so-called Israel matching conditions). Furthermore, considering the mass function of the elastic spacetime
\begin{equation}
m(t,r)=\frac{t}{2}\left(B^3+B-\frac{B'^2}{B}\right),
\end{equation}
equation \eqref{FF01} implies
 \begin{equation}
m^{-} \stackrel{\Sigma}{=}\frac{\alpha^3 R_\Sigma^3 \lambda}{2},
\end{equation}
thus (cf. (\ref{RWm01}))
\begin{equation}
m^{-} \stackrel{\Sigma}{=}m^{+},
\end{equation}
which is also a necessary (but not sufficient) condition in the case of spherical symmetry, as it was first shown in \cite{Fay}.

\subsection{Case: RW with $k=1$ and $a(T)=\alpha T,$ $\alpha>0$}
\label{section2}

Consider as before the junction across a spherically
symmetric surface $\Sigma$ of an elastic non-static spacetime with metric \eqref{est}, that is:
\begin{equation}\nonumber
ds_{-}^{2} = - dt^2 + t^2B^2(r)(dr^2+ d\Omega^2),
\end{equation}
with the Robertson-Walker spacetime
\begin{equation}
ds_{+}^{2} = - dT^2 +\frac{\alpha^2 T^2}{\left(1+\frac{1}{4}R^2\right)^2}\left
( dR^2 + R^2 d\Omega^2\right),\label{RW2}
\end{equation}
corresponding to (\ref{RW1}) with $a(T)=\alpha T$ and $k=1$. The above remarks regarding the meaning and arbitrariness of the signs $-$ and $+$ also apply here.

For the line element \eqref{RW2} one has
\begin{equation}\label{RWrhop}
\rho=3\frac{1+\alpha^2}{\alpha^2 T^2},\;\; p=-\frac{1+\alpha^2}{\alpha^2 T^2}
\end{equation}
and
\begin{equation}\label{RWm}
m(T,R)=\frac{32\alpha TR^3 (1+\alpha^2)}{(4+R^2)^3}.
\end{equation}
Notice that the Dominant Energy Condition is satisfied, and the mass is positive for $\alpha>0.$

We next proceed next as in the previous case; that is, we choose suitable parametrisations for $\Sigma$ in both spacetimes, calculate the corresponding first and second fundamental forms, and demand them to be continuous across the surface $\Sigma$.

The parameters on $\Sigma^-$, and the vector fields spanning its tangent space, as well as the outgoing unit normal vector field, are the same as above, and  the first and second fundamental forms are those given  by \eqref{FF1010} and \eqref{FF2010} respectively.

As for the boundary $\Sigma^+$, the parameters and vector fields spanning its tangent space at each point, can be chosen as in the case above (see \eqref{tangentspaceRW}), whereas the normal vector field is now given by
\begin{equation}
n^{a+}=\frac{1+\frac{1}{4}R^2}{\alpha T}\partial_{R}.
\end{equation}

The first and second fundamental forms for the Robertson-Walker spacetime \eqref{RW2} are now:
\begin{equation}
d\sigma_{+}^{2} =-d\lambda^2+\frac{\alpha^2\lambda^2}{\left(1+\frac{1}{4}R_\Sigma^2\right)^2}R_\Sigma^2 d\Omega^2,
\end{equation}
and
\begin{equation}
H_{11}^{+}=0,\;\;\;H_{22}^{+}=-\frac{4 \alpha \lambda R_\Sigma (R_\Sigma^2-4)}{(4+R_\Sigma^2)^2}.
\label{eq:H}
\end{equation}

Continuity of the first and second fundamental forms implies now
\begin{align}
B\stackrel{\Sigma}{=}\frac{4\alpha R_\Sigma}{4+R_\Sigma^2},\;\;\;B'\stackrel{\Sigma}{=}\frac{4\alpha R_\Sigma(4-R_\Sigma^2)}{(4+R_\Sigma^2)^2}\label{FF}.
\end{align}
where, as before, $\stackrel{\Sigma}{=}$ means that the equalities hold only on the surface $\Sigma$.

Again, as a consequence of the matching, using \eqref{FF}, it follows from \eqref{p1},(\ref{RWrhop}) and (\ref{RWm}) that
\begin{equation}
p_1^{-} \stackrel{\Sigma}{=}p_1^{+}.
\label{FF01a}
\end{equation}
and
\begin{equation}
m^{-} \stackrel{\Sigma}{=}m^{+},
\label{FF01b}
\end{equation}
as it should be expected. In this case, the above magnitudes are:
\begin{equation}
p_1^{-} \stackrel{\Sigma}{=} -\frac{1+\alpha^2}{\alpha^2 \lambda^2},
\end{equation}
and
\begin{equation}
m^{-} \stackrel{\Sigma}{=}\frac{32\alpha \lambda R^3 (1+\alpha^2)}{(4+R_\Sigma^2)^3}.
\end{equation}

\subsection{Case: RW with $k=-1$ and $a(T)=\alpha T,$ where $(\alpha<-1 \text{ and } R>2)\vee (\alpha>1 \text{ and } 0<R<2)$}
\label{section3}

Finally, consider the junction of the non-static elastic  solution (\ref{est}) with the
Robertson-Walker metric
\begin{equation}
ds_{+}^{2} = - dT^2 +\frac{\alpha^2 T^2}{\left(1-\frac{1}{4}R^2\right)^2}\left(dR^2 + R^2 d\Omega^2\right),\label{RW3}
\end{equation}
where $a(T)=\alpha T$ and $k=-1$ in (\ref{RW1}), across a spherically
symmetric surface $\Sigma$.

For the above metric \eqref{RW3} one has
\begin{equation}\label{RWrhop2}
\rho=3\frac{\alpha^2-1}{\alpha^2 T^2},\;\; p=\frac{1-\alpha^2}{\alpha^2 T^2}
\end{equation}
and
\begin{equation}\label{RWm2}
m(T,R)=\frac{32\alpha TR^3 (\alpha^2-1)}{(4-R^2)^3}.
\end{equation}
In this case the dominant energy condition is satisfied and the mass is positive whenever:
\begin{equation}
(\alpha<-1 \text{ and } R>2) \text{ or } (\alpha>1 \text{ and } 0<R<2).
\label{eq:alpha_R}
\end{equation}

The parametrisations on $\Sigma^-$ and $\Sigma^+$ and the orthogonal vector fields spanning their respective tangent spaces at each point are chosen to be the same as in the previous two cases, and so is $n^{a-}$, the normal to $\Sigma^-$, whereas the normal vector field to $\Sigma^+$ is  given by
\begin{equation}
n^{a+}=\frac{1-\frac{1}{4}R^2}{\alpha T}\partial_{R}.
\end{equation}

The first and second fundamental forms for \eqref{RW3} are
\begin{equation}
d\sigma_{+}^{2} =-d\lambda^2+\frac{\alpha^2\lambda^2}{\left(1-\frac{1}{4}R_\Sigma^2\right)^2}R_\Sigma^2 d\Omega^2,
\end{equation}
and
\begin{equation}
H_{11}^{+}=0,\;\;\;H_{22}^{+}=\frac{4 \alpha \lambda R_\Sigma (R_\Sigma^2+4)}{(4-R_\Sigma^2)^2}.
\label{eq:H3}
\end{equation}

Continuity across $\Sigma$ of the first and second fundamental forms implies then
\begin{align}
B\stackrel{\Sigma}{=}\frac{4\alpha R_\Sigma}{4-R_\Sigma^2},\;\;\;B'\stackrel{\Sigma}{=}\frac{4\alpha R_\Sigma(4+R_\Sigma^2)}{(4-R_\Sigma^2)^2}\label{FFII}.
\end{align}
As in the previous two cases, the above equation  \eqref{FFII}, implies that both the radial pressures and mass functions are continuous across $\Sigma$, as  expected.


\section{Elastic thick shells in Robertson-Walker universes}

Next, we will show that the elastic spacetime  metrics (\ref{example1}) and (\ref{example2}) from Example 1 and Example 2, respectively, can be matched to Robertson-Walker spacetimes at the interior and the exterior, leading to models with a well behaved elastic intermediate spacetime, where perturbations propagate in a causal way.

The resulting spacetimes can then be seen as elastic, spherically
symmetric thick shells, separating an interior Robertson-Walker
\lq bubble' from an exterior Robertson-Walker universe with an spatial curvature $k$ different from the one in the inner \lq bubble'.

\subsection{Positive $k$ RW-elastic-flat RW  spacetime}

Consider the metric (\ref{example1}) with $c=1$ and $r_0 =0$. It follows quite
straightforwardly that it can be matched, as an interior solution,
to the flat ($k=0$) Robertson-Walker metric \eqref{RW01} considered
as exterior. To see this, take into account the matching conditions
(\ref{FF01}), which imply $B'\stackrel{\Sigma}{=}B$; it therefore
follows that the  matching radius must be:
\begin{equation}
r_{+}=\frac{1}{2}\ln\left(\frac{1+\sqrt{73}}{6}\right)\approx 0.232.
\end{equation}

Consider next the elastic metric as an exterior solution; it is also easy to show that it can be matched to the $k=1$
Robertson-Walker metric \eqref{RW2} considered as an interior.

To see this, notice that a coordinate change $R=R(r)$ must exist in a neighbourhood of $\Sigma$ such that the matching conditions \eqref{FF} hold at points on $\Sigma$, thus
$$B' = \frac{dB}{dr} = \frac{dB}{dR}\frac{dR}{dr} = \frac{dB}{dR} R'$$
and from the expression for $B$ and $B'$ on $\Sigma$ given by \eqref{FF}, one readily gets from the above equation that $R'=R$, that is:
\begin{equation}
R_\Sigma = \beta e^{r_\Sigma},
\label{eq:aux_1}
\end{equation}
for some constant $\beta>0$. Substituting this into \eqref{FF} we get
\begin{gather}
\frac{\sqrt{3}}{9}(3 \cosh^2 r_\Sigma -1)^{3/2} = \frac{4\alpha\beta e^{r_\Sigma}}{4+\beta^2 e^{2r_\Sigma}}, \label{FFa1}\\
\sqrt{3} \sinh r_\Sigma \cosh r_\Sigma (3 \cosh^2 r_\Sigma -1)^{3/2} = \frac{4\alpha\beta e^{r_\Sigma} (4-\beta^2 e^{2r_\Sigma})}{(4+\beta^2 e^{2r_\Sigma})^2} \label{FFb1},
\end{gather}
which must hold simultaneously, thus producing a system of equations
for the parameters $r_\Sigma$, $\beta$ and $\alpha$. Numerical
calculations show that values of $\alpha$ and $\beta$ exist such
that, for instance $r_\Sigma = 0.100$ is a solution to the above
system, corresponding to $\beta \approx 1.120$ and $\alpha \approx
0.622$.

Thus, the  elastic shell spacetime  is defined for $r\in (r_{-},r_{+})=(0.100,0.232)$ and, as follows from our previous discussions, in this region elastic waves propagate causally (i.e.: $w^2\leq 1$, see Figure \ref{f2}), the metric is regular, and the Dominant Energy Condition is satisfied.


\subsection{Flat RW-elastic-negative $k$ RW spacetime}

We next show that the metric (\ref{example2}) with $r_0 =1/2$ can be matched at the
exterior to the Robertson-Walker metric \eqref{RW3} with $k=-1$ (see subsection \ref{section3} for details).

We proceed in a similar way as in the example above, thus, considerations on the existence of a a coordinate change $R=R(r)$ in a neighbourhood of $\Sigma$, together with (\ref{FFII}), imply
\begin{equation}
R = \beta e^r,\;\;\beta>0,
\label{eq:aux_2}
\end{equation}
on that neighbourhood, which upon substitution into the matching conditions \eqref{FFII} yields
\begin{gather}
\frac{\sqrt{3}}{9}\left[2+3(r_\Sigma - 1/2)^2\right]^{3/2} = \frac{4\alpha\beta e^{r_\Sigma}}{4 - \beta^2 e^{2r_\Sigma}}, \label{FFa2}\\
\sqrt{3} (r_\Sigma - 1/2) \left[2+3(r_\Sigma - 1/2)^2\right]^{1/2} = \frac{4\alpha\beta e^{r_\Sigma} (4+\beta^2 e^{2r_\Sigma})}{(4-\beta^2 e^{2r_\Sigma})^2} \label{FFb2}.
\end{gather}
Again, the above algebraic equations must hold simultaneously for
certain values of the parameters involved: $r_\Sigma, \beta$ and
$\alpha$; thus, for example, $r_\Sigma=0.800$ is obtained for $\beta
\approx 0.264$ and $\alpha \approx 1.024$ hence, the elastic
spacetime can be matched to the exterior Robertson-Walker spacetime
at $r_{+}=0.800$. Note that $\alpha>1$ and $R_\Sigma = 0.588 <2$, as
required (see \eqref{eq:alpha_R}).

At the interior, the elastic spacetime can be matched to the flat ($k=0$)
Robertson-Walker metric \eqref{RW01}: from (\ref{FF01}), and since $B'
\stackrel{\Sigma}{=}B,$ one obtains the matching radius
\begin{equation}
r_{-}=2-\frac{\sqrt{57}}{6}\approx 0.742.
\end{equation}

In this case, the intermediate elastic spacetime is defined for
$r\in (r_{-},r_{+})=(0.742,0.800),$ and in this domain elastic waves
propagate causally (that is: $w^2\leq 1$, see Figure \ref{f3}), the
metric is regular, and the Dominant Energy Condition is satisfied.

\section{Conclusions}

In this paper, we have reviewed some fundamental results on relativistic elastic waves, and have considered in detail the spherically symmetric case, providing an explicit expression for the elasticity tensor $E^{abcd}$ (see\eqref{et}), from where all other relevant tensors (Hadamard elasticity tensor, relativistic Fresnel tensor) can be derived. Further, we have provided an expression for the speed of propagation of the elastic waves $w$, \eqref{w2}, in terms of the constitutive function $v$ and its derivatives (alternatively, in terms of the components of the energy-momentum tensor).

The results thus obtained have been specialized to various cases of spherically symmetric exact solutions previously found by the authors in \cite{BCV}. It is shown that, in all cases but one, there exists an open spacetime region where the solutions
\begin{itemize}
    \item Are singularity free.
    \item Represent elastic matter with a well defined constitutive function $v$.
    \item Satisfy the Dominant Energy Condition.
    \item Are such that the elastic perturbations propagate causally: that is $w^2 \leq 1$.
 \end{itemize}

We also showed that the two non-static elastic solutions discussed can be matched to Robertson-Walker spacetimes, providing two examples of such a matching. These results are then used to build up two spherically symmetric toy-models in which two different Robertson-Walker domains (one flat and one non-flat) are separated by a thick elastic shell (or layer) well-behaved in the above sense, which is well-joined to both Robertson Walker metrics in the sense that the first and second fundamental forms are continuous across the inner and outer surfaces of that shell.

\section*{Acknowledgements}

The authors are grateful to Prof. Ra\"ul Vera, from the Universidad
del Pa\'is Vasco/Euskal Herriko Unibersitatea, for helpful
discussions and suggestions. One of the authors (JC) acknowledges
financial support from the \textit{Spanish Ministerio de Econom\'ia
y Competitividad} through grants REF: FPA2013-41042-P and REF:
FPA2016-76821-P, and also acknowledges the warm hospitality, and
partial financial support, from the University do Minho, where the
present version of this manuscript was prepared. The research of IB
and EV was partially financed by Portuguese Funds through FCT
(Funda\c{c}\~ao para a Ci\^encia e Tecnologia) within the Project
UID/MAT/00013/2013. IB thanks support from FCT, through the Project
PEstOE/MAT/UI0013/2014 and also expresses her thanks and gratitude
for the hospitality at the Universitat de les Illes Balears.

\vspace{1cm}


\end{document}